# From gas to satellitesimals: disk formation and evolution

Angioletta Coradini

*INAF-IFSI*

Via del Fosso del Cavaliere 100, 00133, Rome, Italy

E-mail address: angioletta.coradini@ifsi-roma.inaf.it

Gianfranco Magni

*INAF-IASF*

Via del Fosso del Cavaliere 100, 00133, Rome, Italy

E-mail address: gianfranco.magni@iasf-roma.inaf.it

Diego Turrini

*INAF-IFSI*

Via del Fosso del Cavaliere 100, 00133, Rome, Italy

E-mail address: diego.turrini@ifsi-roma.inaf.it

Abstract

The subject of satellite formation is strictly linked to the one of planetary formation. Giant planets strongly shape the evolution of the circum-planetary disks during their formation and thus, indirectly, influence the initial conditions for the processes governing satellite formation. In order to fully understand the present features of the satellite systems of the giant planets, we need to take into account their formation environments and histories and the role of the different physical parameters. In particular, the pressure and temperature profiles in the circum-planetary nebulae shaped their chemical gradients by allowing the condensation of ices and noble gases. These chemical gradients, in turn, set the composition of the satellitesimals, which represent the building blocks of the present regular satellites.

*Keywords: Solar System: formation; Planets and Satellites: formation; accretion disks; Planets and Satellites: Jupiter; Planets and Satellites: Saturn*

# Introduction

The satellite systems of Jupiter, Saturn and Uranus have been considered as miniature planetary systems since a long time (Greenberg, 1975). While they exhibit very diverse features, their study greatly helped to improve our understanding of the behavior of complex systems formed around a central body in terms of dynamical and physical properties. The regular satellites, which experience negligible solar perturbations, display dynamical properties similar to those of the planets. Moreover, the regular satellites are thought to have formed around their respective planets through a process similar to the one hypothesized for the accretion of the planets around the Sun. Any theory of the origins and evolution of the Solar System must then account for the presence and properties of satellite systems. When studying the origins of the satellites, we cannot avoid putting them in the context of the formation scenarios of their central planets. In fact, it is difficult to imagine satellite formation as totally independent from the environment generated and strongly affected by the central body. However, one has also to take into account that the formation times of the satellites are very short compared to those of their central bodies, and it is therefore possible for several generations of satellites to exist, each formed under different environmental conditions.

# From gas to satellitesimals

If we examine the different satellite systems present in our Solar System we can find both regions characterized by negative outward density gradients and others which seem to have been dominated by stochastic collisional processes. The system of Jupiter exhibits a noticeable regularity. If we plot the so-called reconstructed mass of the satellites (i.e. the mass obtained by adding to each satellite the needed amount of volatile elements, in order to estimate the mass distribution in their formation regions) as a function of the distance from Jupiter (see Figure 1), the resulting density distribution is very regular. The Galilean satellites are a key example of a satellite system which co-formed with its parent planet, the forming satellites accumulating within a circum-planetary accretion disk during the final stages of the planetary formation process. The same doesn't apply to Saturn's system, which shows a large degree of variability (Greenberg, 1975). A large fraction of the mass in the Saturnian system resides in the largest satellite, Titan, which comprises more than 96% of the mass in orbit around the planet. Six other moons, prevailingly of icy composition, constitute roughly 4% of the total mass of the satellite system while the remaining 54 small moons, together with the ring system, only comprise about 0.02%. The surface density computed through the reconstructed mass of the icy satellites (see Figure 1) shows alternating trends, both positive and negative, likely a product of stochastic phenomena at the time of the formation of the system (see e.g. Lissauer, 1995, for a review) As a consequence, it is not straightforward to infer the characteristics of the original disk from the present features of the system like in the case of Jupiter. Moreover, the Saturnian system went subject to an intense bombardment whose presence can be deduced from the overall structure of the icy bodies and from their surface crater distribution. Even more complex is the situation in the satellite system of Uranus, where regular satellites lying on the central plane of Uranus exist, thus orbiting the planet with an inclination of 98° respect to the ecliptic planet. Thus, Uranus and its satellites form a pretty standard system while considered as an isolated dynamical system yet, once considered as part of the Solar System, they represent an anomalous property which must be accounted for by any theory of planetary formation and evolution. Following Greenberg (1975), one can make two hypotheses, depending on the time-scale of the change of the planetary obliquity. The first possibility is that, if the tilting time-scale was long (i.e. the change was gradual), the satellites could have undergone an adiabatic variation of their orbital angular momentum and could have survived the process. The second hypothesis assumes that the change was abrupt (e.g. due to a giant impact of a planetary embryo whose mass was ~7% of that of Uranus, Safronov (1969)). If the tilting took place before the formation of the satellite system, it is plausible to assume that the circum-planetary disk disaggregated and reformed along the new equatorial plane and the satellites then formed in the new disk. If the tilting instead took place after the formation of a satellite system, the original system was probably destroyed and new bodies formed from the debris of the original ones. In this case the link between the satellite system and the formation of Uranus can be weak. Neptune represents another, different case: Goldreich et al. (1989) conjectured that Triton was captured from a heliocentric orbit as the result of a collision with one of Neptune's primordial regular satellites. The immediate post-capture orbit was highly eccentric, with its pericentric distance oscillating periodically. Dissipation due to tides raised by Neptune on Triton caused its orbit to evolve to its present state in $10^9$ years, with Triton being almost entirely molten for most of this time. While its orbit was evolving, Triton cannibalized most of the regular satellites of Neptune and also perturbed Nereid, thus accounting for the satellite's highly eccentric and inclined orbit. The only regular satellites of Neptune that survived were the ones which formed well within 5 Neptune's radii ($R_N$ in the following) and they moved on inclined orbits as the result of chaotic perturbations forced by Triton. Goldreich et al. (1989) calculated that if Triton entered an elliptical orbit with a pericenter at 7 $R_N$ and semimajor axis of $10^3$ $R_N$, then Triton could evolve, through tidal dissipation, to its present situation in less than one billion years. One consequence of the model proposed by Goldreich et al. (1989) was that Neptune would be devoid of satellites between 5 $R_N$ and Triton's current orbit. When Voyager arrived three months later, this was found to be precisely the case. The capture scenario described by Goldreich et al. (1989) was later extended by Banfield & Murray (1992) to take into account the existence and the features of the inner satellites discovered by Voyager. Recently, an alternative scenario was suggested by Agnor & Hamilton (2006) to account for the existence of the outermost irregular satellites. While the implications of the capture of Triton for the Saturnian system will be discussed in detail in the chapter by Mosqueira, Estrada & Turrini, we emphasize that, also in the case of Neptune, the history of the present regular satellites is too complex to trace back the formation process.

Despite the differences previously described, Canup and Ward (2006) noted that the ratios of the overall masses of the regular satellite systems to those of their parent planets are about $10^{-4}$ for the systems of Jupiter, Saturn and Uranus. This suggests that a common process took place in all the satellite systems and in particular that the mass fraction of the satellite systems is regulated by the balance in the disk between the supply of inflowing material and the satellite losses due to orbital decay driven by the interaction with the surrounding nebula. As a consequence, several satellite generations could have formed and got lost toward the central body, until the mass inflow onto the disk ended. The present satellites systems are likely the surviving members of the last satellite generation. However, in their work, the authors did not relate the mass accretion rate and the disk features to the formation time-scale and characteristics of the central body, so that it is almost impossible to date the formation time of the satellites.

Irregular satellites, instead, are believed to have formed in the Solar Nebula and to have been gravitationally trapped by the giant planets at a later time. Irregular satellites can supply information on the conditions of the surrounding space, as well as on the characteristics of the planets at the time when their trapping efficiencies were higher. The details and implications of the capture of the irregular satellites will be discussed in the chapter by Mosqueira, Estrada & Turrini. A complete and exhaustive model of satellite formation shall take into account the conditions existing in the Solar Nebula, since they represents the boundary conditions to be considered. However, connecting a model of the Solar Nebula to a model of giant planet formation that includes the formation of regular satellites is quite a difficult problem. Due to the strong link existing between the formation of the disk and that of the central body, we cannot avoid mentioning the current theories on the formation of giant planets.

In the case of Jupiter and Saturn, the study of the formation conditions of the central bodies can be more directly linked to those of their surrounding disks, whose thermodynamic conditions, in turn, represent the boundary conditions for satellite formation and thus determine the initial chemistry of the satellites. Jupiter and Saturn are mostly gaseous and likely accreted a large part of their masses from the proto-Solar Nebula when the gas was still abundant. The state of the art on this subject has been reviewed by several authors (see e.g. Lunine et al., 2004; Lissauer & Stevenson, 2007). The formation of giant planets is of paramount importance not only in order to understand our Solar System, but also to shed light on the possible nature and evolution of extraterrestrial planets. Moreover, following Guillot et al. (2004), we can state that Jupiter, owing to its large mass and rapid formation, played a crucial role in shaping the Solar System as we know it today. Jupiter mostly contains hydrogen and helium (more than 87% by mass) and as such it bears a close resemblance to the Sun. However, the Sun has only 2% of its mass in heavy elements whereas Jupiter has between 3-13%. Spectral observations of Saturn from the far infrared spectrometer aboard the Cassini spacecraft revealed that the C/H ratio in the planet is in fact about twice higher than previously derived from ground-based observations and in agreement with the C/H value derived from Voyager IRIS (for a discussion see Hersant et al. 2008.). The exact amount of these heavy elements in the planet and their distribution are key issues to understand how the Solar System formed (ibid), and the only way to correctly evaluate them is to know what the mass of the planetary core is. In fact, the idea that the atmospheres of the giant planets were similar in composition to the primordial nebula from which they formed has been largely disproved thanks to their study both from space and from Earth (Encrenaz 1994). Encrenaz (1994) noticed that significant differences were measured in the abundances of helium, deuterium and carbon of the four giant planets, both between themselves and in respect to the Solar Nebula. In particular, the variations in the C/H and D/H ratios gave support to the "nucleation" formation scenario, where the four giant planets first accreted a nucleus of about 10 Earth masses, big enough to gravitationally capture the surrounding gaseous nebula (ibid). One of the problems still open in this scenario is the mass of the central core that should be considered. Goldreich et al (2004) revised the problem of initial formation of planets and suggested that the region populated by the outer planets could have been characterized by the presence of many large bodies, about the size of Uranus and Neptune, which could have formed earlier than the inner planets provided that the outer disk was richer in mass than the inner one. This means that the original disk from which the planets evolved contained the mass of a "minimum mass nebula" in the inner regions, but a much larger amount in the outer ones. Therefore, the core of Jupiter and Saturn could have formed at this time, much before the time when the nebular gas dissipated.

Another important consideration on the possibility to quickly form the central core of giant planets has been done by Kornet, Wolf & Rozyczka (2007), who showed that the frequently adopted approximation that the surface density of the planetesimals is a constant fraction of the gas surface density cannot be accepted without care. In fact, these authors showed that the surface density of planetesimals can be locally enhanced by a large factor respect to the gas, thus speeding up the subsequent formation of the planets. This is due to the interaction between the gas and the solid particles in the initial phases of the disk evolution, as well as to the mutual interactions between the growing planetesimals. Therefore, a detailed analysis of the initial conditions present in the Solar Nebula could help solving the problem of the formation of the cores of the giant planets.

## Gas capture and the formation of disks and planets

The formation of Jupiter and Saturn can be modeled through the gas accretion onto a pre-existing solid core. This approach, which has been adopted by several authors, can be followed in all its different phases, provided that a detailed description of the gas accretion process and the correct evolution of the central body thermodynamics are considered. In fact, the central body evolution is not irrelevant for the thermal evolution of the disk surrounding it, and, in turn, the hydrodynamic evolution of the disk is responsible for the mass accretion rate onto the planet, which at this stage is the main energy source of the growing object.

For the rest of this section we will consider mainly the cases of Jupiter and Saturn, for which several authors make the following working hypotheses. First, a scenario in which the regular satellites form within circum-planetary accretion disks produced during the final stages of gas accretion (e.g. Lubow & Ogilvie (2001); D'Angelo, Henning & Kley (2002); Magni & Coradini (2004)) is assumed. Second, for a given inflow rate of gas and solids, the circum-planetary gaseous disk is assumed to be in a quasi-steady state due to the balance of the inflow supply and the disk's internal viscous evolution, assuming that the disk viscous spreading time is short compared to the timescale over which the inflow changes. Up to now, the existing hydrodynamic models generally used spherically symmetric proto-planets, consisting of a growing solid core and a solar composition gaseous envelope, surrounded by the accretion region. The accretion of giant planets, however, can be hardly assumed as spherically symmetric since it takes place in a strongly asymmetric region of several AU, dominated by the gravity of the Sun and crossed by density waves. On the contrary, three-dimensional models of the evolution of giant solar and extrasolar planets didn't include satisfactory physical description of the interfaces between the growing planet and the nebula (e.g. Artymowicz (1998); Rauer & Hatzes (2007) and references therein). Lissauer et al. (2009) consider the gas accretion in a turbulent disk, whose evolution is computed in a three-dimensional space. This allows these authors to improve their modeling of Jupiter's formation, taking into account the time evolution of the gas flux throughout the Hill sphere. Their modeling, however, is based on ad hoc assumptions on the gas capture criterion. Their results suggest accretion times for Jupiter's gaseous envelope from the circum-Jovian disk ranging between several $10^4$ years to a few $10^5$ years, the accretion timescale strongly depending on the chosen physical parameterization and the disk turbulence. Moreover, while the influence of the gas accretion onto the planet is correctly taken into account, the feedback effects of planetary formation on the disk are not explicitly considered (ibid).

The physical nature and the development of the core instability need to be discussed more deeply. First we have to introduce the concept of *critical mass*, i.e. the mass of the core beyond which the core starts collecting the surrounding gas. Theoretical works have modeled the process of gas accretion invoking hydrodynamic instability (Magni & Coradini, 2004) or quasi-hydrostatic, irradiation-driven compression (Lissauer et al., 2009). At present, the observational constrains are too limited to discriminate between the two mechanisms. Analytical arguments (Stevenson, 1982; Wuchterl, 1993) and numerical calculations (Mizuno, 1980) suggest that the onset of core instability occurs when the mass of the gaseous atmosphere around the core becomes comparable to the mass of the core itself. In a recent paper, Rafikov (2006) discussed the subject, determining the value of the critical mass in different conditions. This value depends on the physical characteristic of the atmosphere of the accreting object and mostly on the gas opacity. Rafikov (2006) systematically studied quasi-static atmospheres of accreting protoplanetary cores for different opacity behaviors and realistic planetesimal accretion rates in various parts of the protoplanetary nebula. In the region of the giant planets, he considered values of the critical mass as high as 20-60 Earth masses (for opacity values in the range 0.1-1 $cm^2g^{-1}$) if planetesimal accretion is fast enough for protoplanetary cores to form prior to the dissipation of the nebular gas. If the mass of the forming planet is much less than the critical value, the envelope/core mass ratio is very low. As a consequence, the self-gravitation of the gas is negligible and the maximum equilibrium radius of the envelope is much larger than the Hill's radius. Therefore, there is a limit to the amount of gas that can be collected by the planet from the surrounding nebula. When the mass of the forming planet is near to the critical value, also the envelope radius is comparable to the Hill's radius (about a quarter of the Hill's radius following Lissauer et al., 2009). For proto-planetary masses greater than the critical value, a further increase produces a growth of the Hill's radius greater than the corresponding value of the envelope radius, and more and more gas can be collected from the nebula. At the same time, the heating due to the in-falling gas produces an expansion of the envelope up to the boundaries of the Hill's lobe. The gas accretion is thus driven by the thermodynamic and hydrodynamic conditions near the growing planet and, in turn, depends on the ratio between the energy carried by the in-falling gas and the energy associated with pressure effects. In addition, angular momentum is redistributed in the in-falling gas under the action of the gravitational fields of the Sun and of the growing planet. Moreover, fast radial hydro-dynamical instabilities can grow until they stop the accretion, producing a rebound of a significant fraction of the gas envelope out of the sphere of influence of the growing planet, as shown by Wuchterl (1991). Such effect, however, strongly depends on the assumed boundary conditions, particularly on the density and temperature of the gas in the Solar Nebula (Wuchterl, 1993) and on the adopted geometry. As a result of these combined effects, the accretion timescale is the maximum between the radiative timescale and the timescale needed by the external regions of the envelope to reach a hydrostatic quasi-equilibrium state. However, dissipative processes due to turbulence are present in the Solar Nebula and in particular around the growing planet. Here, angular momentum is dissipated inside the accretion disk formed by the in-falling gas producing a substantial flux of matter towards the core of the protoplanet, strongly shortening the accretion timescale. Thus, the final gaseous accretion of the central planet occurs via formation and replenishment of a disk-shaped envelope embedded in the Hill's lobe, which suggests again that the rapid collapse of a spherical gaseous envelope is not a valid description of the process. A 3D hydrodynamics code must be used to trace the accretion. In the next section we'll describe our three dimensional model of planetary accretion in some detail to supply a template of the relevant processes and their treatment.

## Evolution of the circum-planetary disk: a template model

A few years ago Stevenson (2001) discussed the formation of the Galilean moons, which was thought to take place with a mechanism similar to that of planetary formation in a disk of gas and dust swirling around the proto-Jupiter. As Stevenson pointed out, this mechanism works only if the collapse time of the growing planet is shorter than the accretion time on the planet of the forming satellites, and these constraints aren't respected in the Jovian system (ibid). Stevenson noticed that Callisto's composition could be an indication that the satellites were formed at the end of Jupiter's gas accretion, when the gas density in the disk was tenuous. Several authors attempted to quantitatively approach this problem: here we report some of the most important results obtained up to now.

In what follows we will describe the structure and the results of different models of the evolution of circum-planetary disks, in order to show what are the critical parameters affecting the satellite formation process. To do that, we have to set some preliminary definitions. We will call *growing planet* the central body in nearly hydrostatic equilibrium, containing a planetary core composed by a mixture of rocky and icy materials plus a spherical gaseous envelope. Surrounding the growing planet is a deeply anisotropic region where gas accretion is supersonic and turbulence is present: this region will be referred to as the *accretion disk* (Shakura & Sunyaev, 1973). Outside the accretion disk is the *Hill's lobe*, the large spheroid region where gas accretion takes place and through which material captured from the Solar Nebula evolves inward. These three regions represent a useful schematization of the history of the gas which enters the sphere of influence of the growing planet. The accretion on the planet can be treated assuming an annular region, centered on the Sun, as the planet-feeding zone. The radial extension of the feeding zone should be large enough to include the region where the gravitational and dynamical effects (i.e. the density waves induced by the planet on the gas of the Solar Nebula) due to the growing planet are significant.

Alibert et al (2005) modeled the Jovian disk following the time evolution of its surface density in steady conditions. They also try to take into account of the local vertical disk structure by solving the barometric equation. The disk is fed by the gas captured from the protosolar nebula. The thermal effects of radiation emitted by the growing planet on the disk structure are neglected, as well as the interactions between the large atmosphere surrounding the proto-Jupiter and the disk. Once the protosolar nebula is dissipated, the disk is not fed anymore while it keeps losing mass toward Jupiter. The associated angular momentum transfer causes an outward expansion of the disk. The effects of this expansion aren't explained in the paper (ibid). The two phases considered are totally decoupled and it is not clear how they are related to the overall evolution of the central planet. The disk is turbulent and the strength of the turbulence is modeled through the parameter α, which is treated as a free parameter. The external radius of the sub-disk is treated as well as a free parameter. Owing to these assumptions, the authors show that the satellites can survive both Type I and II migration effects (Ward, 1997). Even when giving a good phenomenological description of the involved processes, this approach is however non-predictive, since the parameters describing the disk are poorly constrained. Moreover, the authors reduce the effect of Type I migration by modifying it through a suitable factor *f* that can vary between 1- 0.001. The authors modify the values of α until the temperature of the disk allows the formation of an object with the same rock/ice ratio assumed for Callisto, while the low value of *f* permits its survival to type I migration. This method can supply some insight in the thermodynamical conditions of the satellite disk, but does not permit to relate them neither to the overall evolution of Jupiter, nor to the mass present in the disk, which in turn is a function of *f*. This evolutionary model has been used in order to compute, in a following paper (Mousis and Alibert, 2006) the detailed thermodynamical conditions of the disk, as well as the associated condensation sequence. In conclusion, this model could be a useful first order evaluation of the thermodynamic conditions of the disk. However, the large number of unknown parameters present in the work does not allow using it in a predictive way.

Another approach is the one suggested by Canup and Ward (2002), based on the idea that the disk formed at the end of the gas accretion and therefore was gas-poor. The main argument of these authors is that a massive disk would be characterized by too high temperatures to allow for ice condensation in the region of Ganymede and Callisto. Such consideration holds true unless the disk has very low viscosity ($\alpha \sim 10^{-6}$), with a correspondingly long disk viscous lifetime of $\sim 10^{6}$ yr. In massive disks, the lifetimes of Galilean-sized satellites against Type I orbital decay due to disk torques and aerodynamic gas drag could be extremely short ($10^{2}$ and $10^{3}$ yr respectively). Due to Type II orbital decay, also larger satellites can be lost on short timescales if the satellites do not succeed in opening gaps in the disk. This implies that only satellites that are large enough to open a gap can survive. The model by Canup and Ward (2002) is the one that better describes the present configuration of the Galilean satellites, yet it is not clear when such "starved disk" formed across the formation process of Jupiter. Moreover, the model does not satisfactory explain the weak density gradient presently measured in Jupiter's system.

Following Ruskol (2006), who reports the results obtained by the Russian school on this subject, it is very difficult to decide whether large mass or small mass disks should be favored. Makalkin et al. (1999) explored the evolution of a large mass disk, similar to the one described in Coradini and Magni (1984). In a later work, Makalkin and Ruskol (2003) investigated the evolution of a low mass disk. In their modeling, the authors used an α model to describe the disk, assuming a value of $\alpha \sim 10^{-3}$. In both cases the authors were able to reconcile the disk thermodynamic conditions with the chemical composition of Galilean satellites. This ambiguity is substantially due to the fact that the assumptions on the disk derive by the present observations on the satellites. Therefore, as in any inversion problem, there are several parameters that can be adjusted and tuned. In particular, the mass accretion rate and the strength of the viscosity, as defined through the parameter α, are strongly model-dependent. A more robust approach could be to relate the characteristics of the disk to the evolution of Jupiter, as we describe here following the approach used by Coradini and Magni (2004).

Let then consider a forming proto-planet collecting gas and dust from the protosolar nebula. The reference system is rigidly rotating with an angular velocity equal to the Keplerian orbital velocity at the location of the planet. The numerical integration of the system of partial differential equations is performed by using a first order explicit numerical scheme, the so called *donor cells* method described by Boss (1980). An advantage of this method is that it ensures rigorous conservation of the global mass and momentum through the numerical grid. Moreover, in those regions of severe mass depletion, the method prevents negative densities by taking sufficiently small time steps. Note that tidal and Coriolis forces are present, since the grid is rotating with the growing planet. With this scheme we can properly account for the mass and angular momentum transfer from the Solar Nebula to the planet. However, due to the discrete structure of the hydrodynamics equations, angular momentum is no longer exactly conserved as it would be in an inertial, non rotating reference frame. The shorter is the timescale of the spurious diffusion, the more the angular momentum conservation is violated. The timescale of the spurious diffusion can be increased by increasing the mesh grid-points number, at the cost of increasing the computational time: how to achieve a balance between these two competing needs has to be evaluated depending on the aims of the simulations. The grid covers a shell around the present orbit of the planet which has more elements in the volume surrounding the protoplanet itself (i.e. accretion zone) than elsewhere around the orbit. The angular vertical thickness of the feeding zone is 10°, i.e. large enough to contain more than one pressure height-scale. Due to the reasons previously presented, the radial extension of the feeding zone has been assumed ranging from 2.7 AU to 13.7 AU.

The hydrodynamic evolution is driven by the ordinary mass, momentum, energy conservation and radiative diffusion equations. The external forces are the pressure forces, plus non-inertial forces, i.e. Coriolis and centrifugal, and gravity forces due to the Sun and to the growing planet. In addition, the numerical model also includes a first order treatment of the self-gravitation of the gas, which may be non-negligible inside the Hill's lobe and near the growing planet due to the strong density enhancement. The self-gravity term is computed starting from the radially averaged density around the growing planet. The equations are expressed in spherical coordinates and in the Eulerian approach.

The effects of turbulent viscosity on the gas are directly treated into the hydrodynamic equations of momenta adding the ordinary nonlinear terms (Landau & Lifschitz, 1971) like pseudo-acceleration terms. The main parameter involved in the turbulent terms is the so-called α parameter (Shakura and Sunyaev,1963) that tunes the strength of the turbulence itself. Turbulent dissipation causes a general outward transfer of angular momentum in the nebula and an inward flux of matter. In the region around the growing planet, the geometry of the forces causes respectively outward and inward fluxes of angular momentum and mass. So, turbulent viscosity causes an enhancement of the mass accretion rate of the growing planet. Since around the growing planet the adopted grid geometry contrasts with the natural planet-centered geometry of the fluxes, we considered averaged values of the radial fluxes. The value of α has been assumed to be 0.05 in the case of Jupiter, and 0.01 in the case of Saturn. This has been done since the examination of the structure of the Jovian disk exhibits a much higher degree of turbulence then the Saturnian one. The accretion timescale for Saturn has been found to be from two to four times that of Jupiter. In this calculation, the accretion timescales for Jupiter and Saturn vary between a few $10^3$ and about $10^5$ years, depending on the value of the α parameter. In Figure 2 we plotted the density distribution on the mid-plane for the accretion regions around Jupiter and Saturn respectively. The two strongly anisotropic patterns show the regions where the accreting gas is concentrated before entering the envelope of the growing planet. The central region is nearly isotropic due to the combined effects of the non inertial forces, the pressure and the viscosity, which cause the rotation to change from retrograde to prograde. Therefore, in the region close to the planet a Keplerian, gravitationally bound prograde disk can be identified. The region from which the protoplanet collects the gas, i.e. the feeding zone, is shown in Figure 3. The feeding zones of Jupiter and Saturn do not overlap between themselves, thus allowing treating the formation of the two planets independently. The thermal evolution of the disks has been evaluated integrating the total energy equation and including the radiative time-dependent diffusive terms.

## The disk evolution and chemistry

As already stated, the gas is captured by the planet either directly or through the formation of an accretion disk. The actual mechanism depends on the distribution of the angular momentum of the gas. The structure of the growing planet, and, in particular its mass, radius and luminosity, depends on the rate of mass accretion. The distribution of the gas swirling around the planet depends, in turn, on the planet's thermodynamic characteristics. A fraction of the energy collected by the planet through mass capture is in fact released back to the disk as thermal energy. The region surrounding the growing planet is heated up by irradiation, the pressure in the region increases, and a pressure barrier reduces the mass flow (Magni and Coradini, 2004). This is a negative feedback effect that tends to increase the planet growth timescales and affects the overall disk thermodynamics.

The energy production rate per unit mass is assumed to be constant throughout the planet and the energy acquired by the planet is virialized. The energy flux rate can be computed simply as the ratio of the total energy flux per unit mass that enters the region occupied by the planet. The opacity coefficient has been computed by combining the results of different authors: we refer the readers to Magni & Coradini (2004) for details on the opacity treatment. The equation of state, linking the pressure to the other thermodynamic variables, includes both partial dissociation of $H_2$, ionization of H and He, as well as electronic degeneracy that can become important in the inner part of growing planet. The dimension, the structure and the global parameters of the circum-planetary accretion disk are evaluated basing on the angular momentum distribution (referred to the growing planet) of the cells inside the Hill's lobe.

In order to supply realistic information on the accretion timescales, the mass distribution and the thermal evolution of the region where satellite formation takes place, the authors considered a forming planet embedded in a *minimum mass* Solar Nebula (i.e. about 0.02 solar masses). The growth of the planet is followed from the initial, critical mass (i.e. about 15 Earth masses) up to the final planetary mass. The mass of the feeding zone has been artificially depleted after the planet reached its final mass in order to simulate the natural process of depletion due to gap formation and/or the process of dissipation of the nebula driven by the solar wind. We obtained a sequence of quasi-stationary models, which are characterized by decreasing masses and optical depths, of the formation region of the satellites. Basing on these results, we studied the saturation pressures of the main chemical volatile components, like $H_2O$, CO, $CO_2$, $CH_4$, $NH_3$, Ar, Ne. In Figure 5 we show the temperature distribution in the disks around Jupiter and Saturn at different phases of the accretion process of their respective planets. As we previously said, these two disks are characterized by different values of the average α parameter: in the case of Jupiter ($\alpha$=0.05), the disk is more turbulent then in the case of Saturn ($\alpha$=0.01).

The final phases of accretion here described are slow: the disk gradually cools down, allowing for the condensation of different materials. To follow these phases in a quantitative way, we computed the evolution of the temperature in the disk. These calculations take into account the average variation of the temperature, since different parts of the disk can be characterized locally by different values of the thermodynamic quantities. The structure of the disk is in fact extremely complex: the external region is strongly perturbed by the interactions with the gas of the Solar Nebula, while the internal regions are dominated by the interaction with the protoplanet.

The temperature of the Jovian disk keeps almost constant for a large fraction of its history in all the regions in which we have divided the disk for sake of simplicity. These regions correspond respectively to 0-40 $R_J$, 40-50 $R_J$, 60-80 $R_J$, 80-100$R_J$ and 100-150 $R_J$. Obviously, the inner region (0-40 $R_J$, i.e. the one which is relevant for satellite formation) is the hottest one. Toward the end of the accretion process, when the mass inflow reduces, the disk cools down. At this stage, the Jovian disk reaches about 125°K in the region between 0-40 $R_J$, and 68°K in the region between 80-100 $R_J$. The Saturnian disk is divided in the same number of regions and is characterized by a similar behavior, but the disk is generally colder.

At the end of the accretion phase, the temperature becomes quite uniform in the entire disk, stabilizing at about 39°K.

The gas pressure in the disks can be computed also as a function of the accretion time: the results are shown in Figure 6. As the temperature, the global gas pressure and the partial pressures of the different gases decrease as the mass inflow slows down. The mean values of the ratios between the disk pressure and the saturation pressure of the different gases present in the disk are shown in the same Figure. When the ratio is greater than unity, the component is in gaseous form; when it is lower, the thermodynamic conditions in the disk allow for a phase transition from gas to solid. It is obvious that this isn't a complete chemical evolution model, since we don't compute exactly which are the species that are present in the disk at each time step. Our calculations are indication that, provided certain species are present in the disk, the thermodynamic conditions could allow their condensation. We plan to improve such evaluation in the near future by developing a more complete model of the gas chemistry, which would use as its input the results of the hydrodynamical model. From a qualitative point of view, therefore, in Jupiter's case (Figure 6, panels a-c)only $H_2O$ condensation is possible in the formation region of the satellites (0-40 $R_J$) and only in the final phases of the accretion. When the accretion stops, other species start to condense (e.g. $NH_3$ and $CO_2$ ices) both in the inner and outer regions. The emerging picture is very interesting since it hints that a certain amount of ammonia should have been incorporated in the satellites, possibly in the form of ice or in the more complex molecules of clathrates, provided that the relative abundances of N and H could allow their formation. The presence of ammonia is important since it significantly lower the freezing point of the material in which it has been incorporated, therefore influencing the thermal history of the satellite. Our analysis implies that the Jovian disk was dominated since the beginning of its history by the presence of $H_2O$ grains. Silicate particles were also present in the Jovian disk. Icy particles contributed, with the silicate grains, to the formation of *satellitesimals* in the disk: the combination in different proportions of these two main components played the most important role in the determination of the average density of the satellites.

In the case of Saturn the lower temperatures allow for the condensation of $CO_2$, $NH_3$ and $CH_4$ grains is possible (see Figure 6, panels d-f). At this stage, in fact, most of the Saturnian disk is as cold as the surrounding solar nebula from which the gas is collected. Only the internal regions of the disk are hotter, having temperatures which cause the sublimation of different ices, and thus permit a complex chemistry to evolve. Highly volatile gases can be incorporated in the mixture since the very beginning of the life of the disk. Even Ar can be incorporated in the satellitesimals in the external part of the disk, in agreement with the puzzling features of the atmosphere of Titan, where no primordial noble gases other than Ar were detected by the Gas Chromatograph Mass Spectrometer (GCMS) aboard the Huygens probe during its descent to Titan's surface on January 14, 2005. Other gases were perceptually below the detection level of GCMS (Kr and Xe). The CO ice is not condensed in the Saturnian sub-nebula. As a consequence, the satellitesimals formed in the disk could be depleted of this ice as suggested by Mousis et al. (2008). This would explain the deficiency of CO observed in Titan's atmosphere (Alibert & Mousis, 2007). $H_2O$, $NH_3$, $CO_2$, and $CH_4$ would instead remain incorporated in solids. The complex chemistry present in the Saturnian disk can thus be responsible for the internal evolution and differentiation of the satellites.

On the surfaces of the icy satellites, high volatility ices could be lost during the further evolution of these bodies: the icy moons, in fact, were intensively bombarded and underwent strong interactions with both solar and planetary radiations, as well as by high energy electrons and protons trapped in the radiation belts of Saturn. Only Titan has preserved a large part of the original chemistry.

Water ice is abundant on Saturnian satellites, as shown by the near-infrared spectra returned by VIMS (see chapter by Dalton et al., this book). The ice phase can be used to study the crystalline order of the surface ice on all the satellites. It could also be a way to measure the lattice order of ice that, in turn, depends on its condensation temperature and rate, its temperature history, and its radiation environment. Inner satellites are characterized by a larger content of crystalline ice, in respect to the amorphous one (which is almost absent), as measured on the basis of the strength of the absorption bands at 2 and 2.05 μm.

The role of carbon compounds was also important for the overall chemistry together with the water ice particles. Satellitesimals formed in the Jovian and the Saturnian disks would probably maintain records of the original composition. A similar analysis performed on the $CO_2$ ice seen by VIMS shows that a large amount of "organic contaminants" was present in the Saturnian system. The radial distribution of the "contaminants" across the Saturnian system can be measured through the visible spectral slopes (ibid). The satellites have red (positive) slopes in this range, probably caused by the presence of a still unidentified UV organic absorber. Therefore, we can argue that the original chemistry is not totally cancelled by the overall evolution of the satellites.

At this point of the evolution of the disks the model developed by Canup & Ward (2002) becomes important. In fact, the end of the accretion of the planet is accompanied by the formation of a tenuous disk like the one described by these authors. For this reason, the previously presented results aren't in disagreement with their modeling: on the contrary, they confirm their assumptions in a rigorous way. The further evolution of the disk could imply displacements of the satellitesimals as well as capture and bombardment from external bodies. These subjects will be discussed in the second part of this chapter (see Mosqueira, Estrada and Turrini, this book).

# Figures

## *Figure 1*

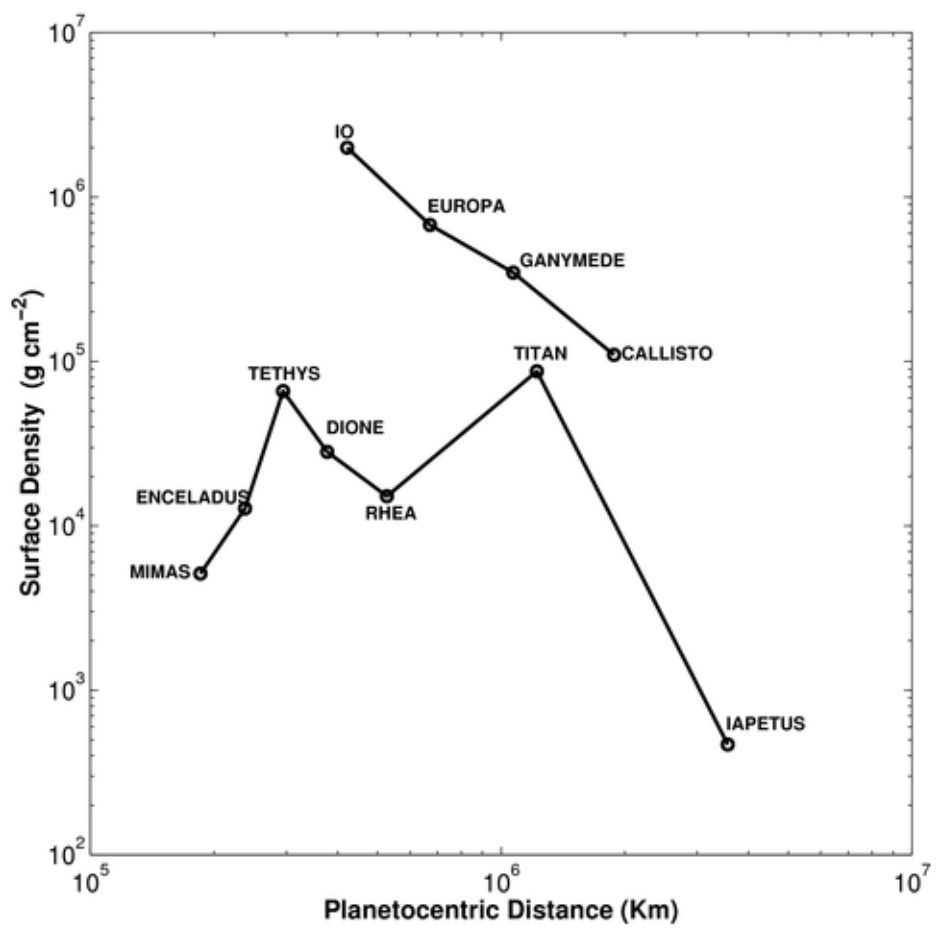

Figure 1: Surface density of the circum-planetary disks of Jupiter and Saturn estimated from the reconstructed masses of the satellite systems. The reconstructed mass is obtained by adding to each satellite the needed amount of volatile elements to achieve solar composition. Surface density is expressed in g cm$^{-2}$ while radial distance from the central planet is expressed in Km. Figure adapted from Pollack & Consolmagno (1984). The protosatellite disk surface density profile was constructed by spreading the mass of each satellite, augmented to match a solar proportion of elements, over an annulus centered on the current satellite orbit. Similarly dense disks have been used in previous models of Galilean satellite formation.

*Figure 2*

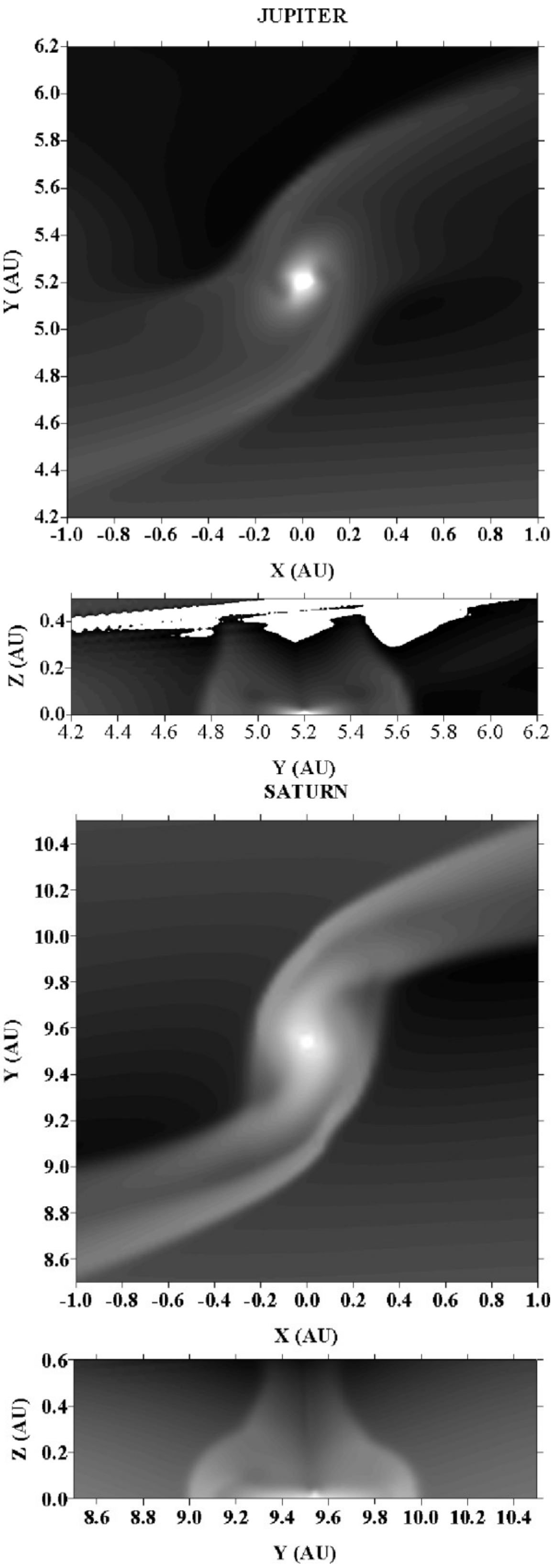

Figure 2: Density distributions in the accretion regions of the growing planets during the final stages of their formation. For each planet, the upper plot represents the mid-plane of the Solar Nebula while the lower plot shows the vertical section along the Sun-planet direction. All distances are expressed in astronomical units while density is expressed in g $cm^{-3}$ and in logarithmic scale.

*Figure 3*

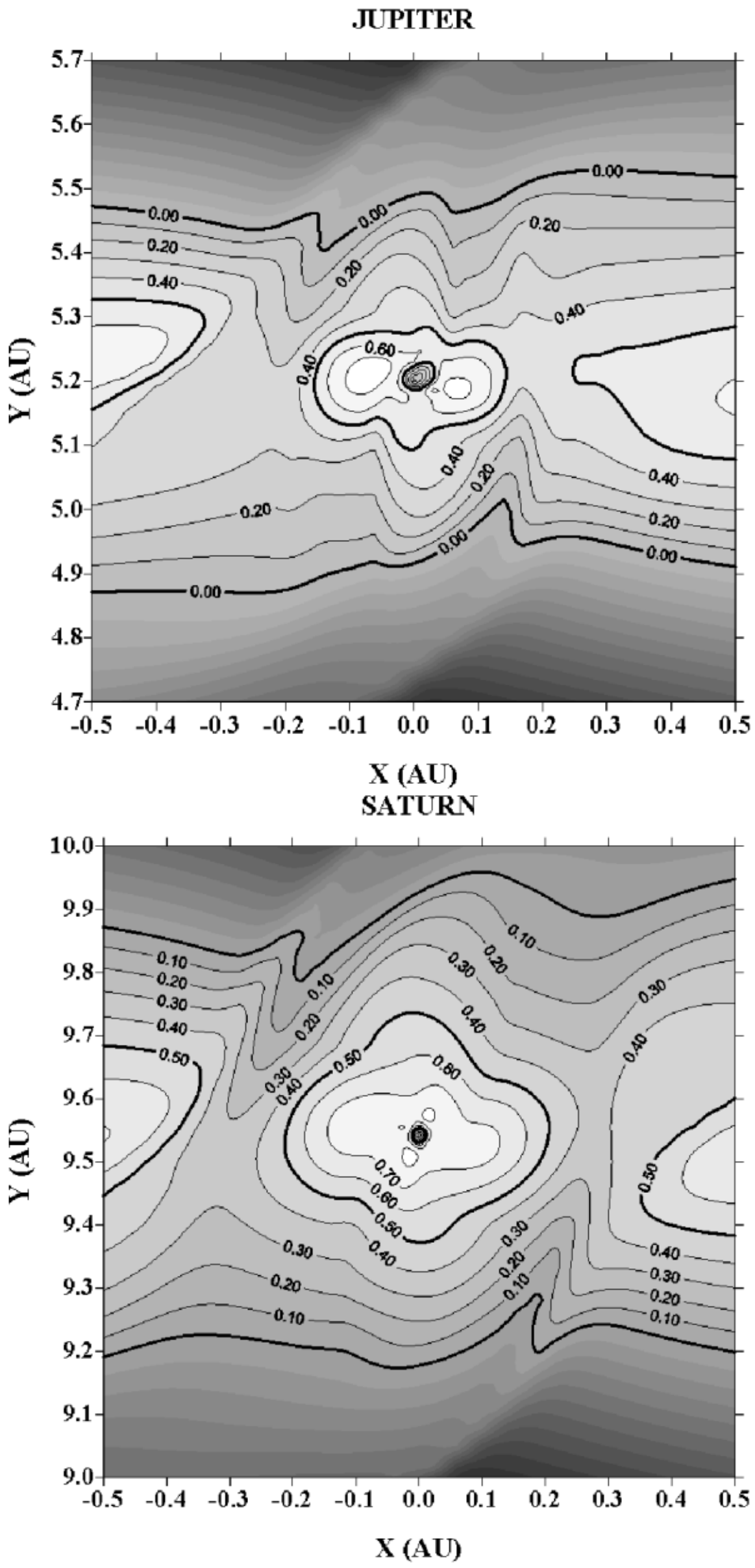

Figure 3: Mapping of the tangential velocities of the in-falling gas for Jupiter and Saturn at the same evolutionary stage as Figure 2. Tangential velocities are expressed in units of the local circular Keplerian velocity while distances are expressed in astronomical units. The iso-level curves refer to the regions of prograde motion around the planet. The Jovian Hill's radius is about 0.35 AU while the Saturnian one is about 0.43 AU.

*Figure 4*

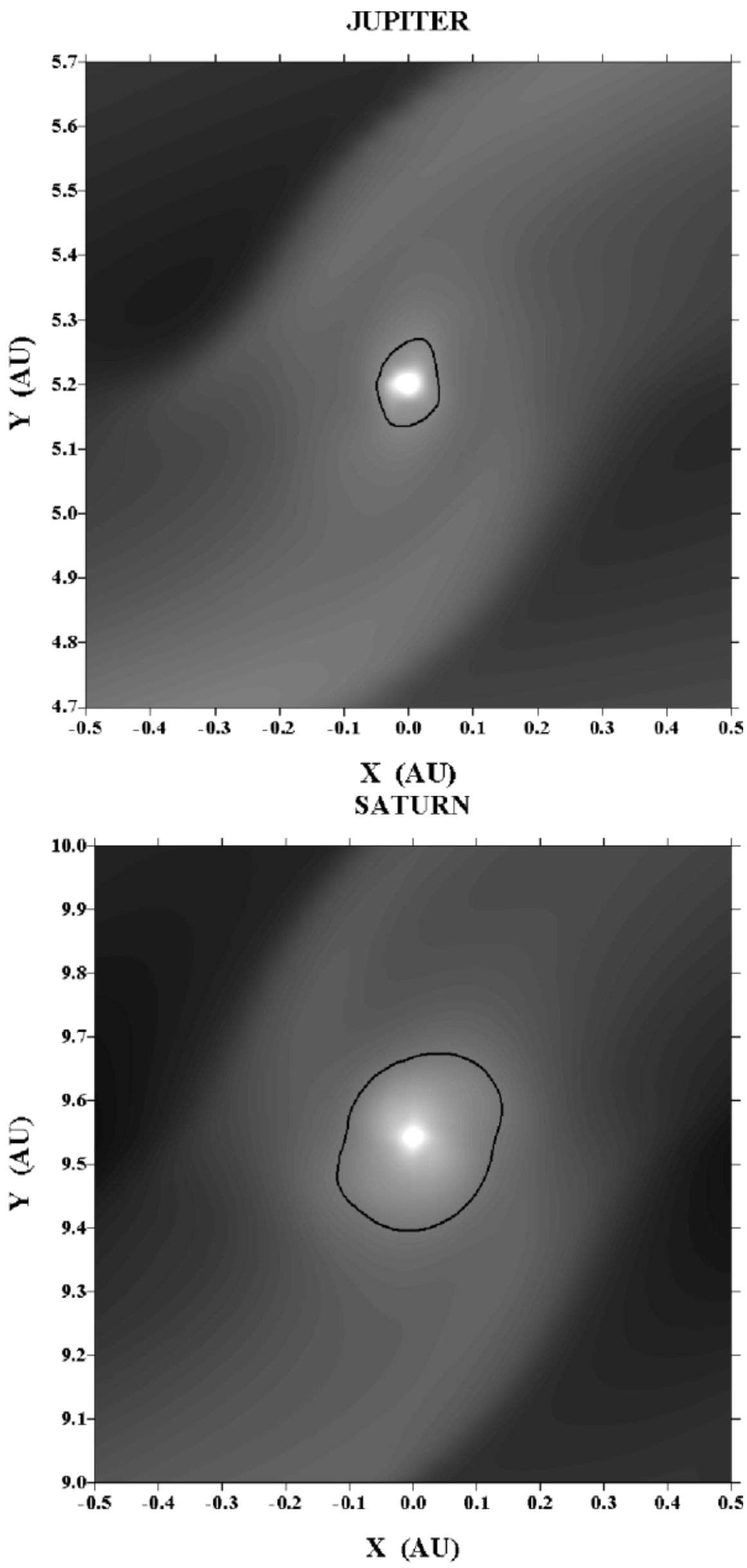

Figure 4: Surface density distribution for Jupiter and Saturn at the same evolutionary stage as Figure 2. The black iso-curve refers to a density value of 1/30 of the central density. The surface density is expressed in logarithmic scale in cgs units while distances are expressed in astronomical units.

*Figure 5*

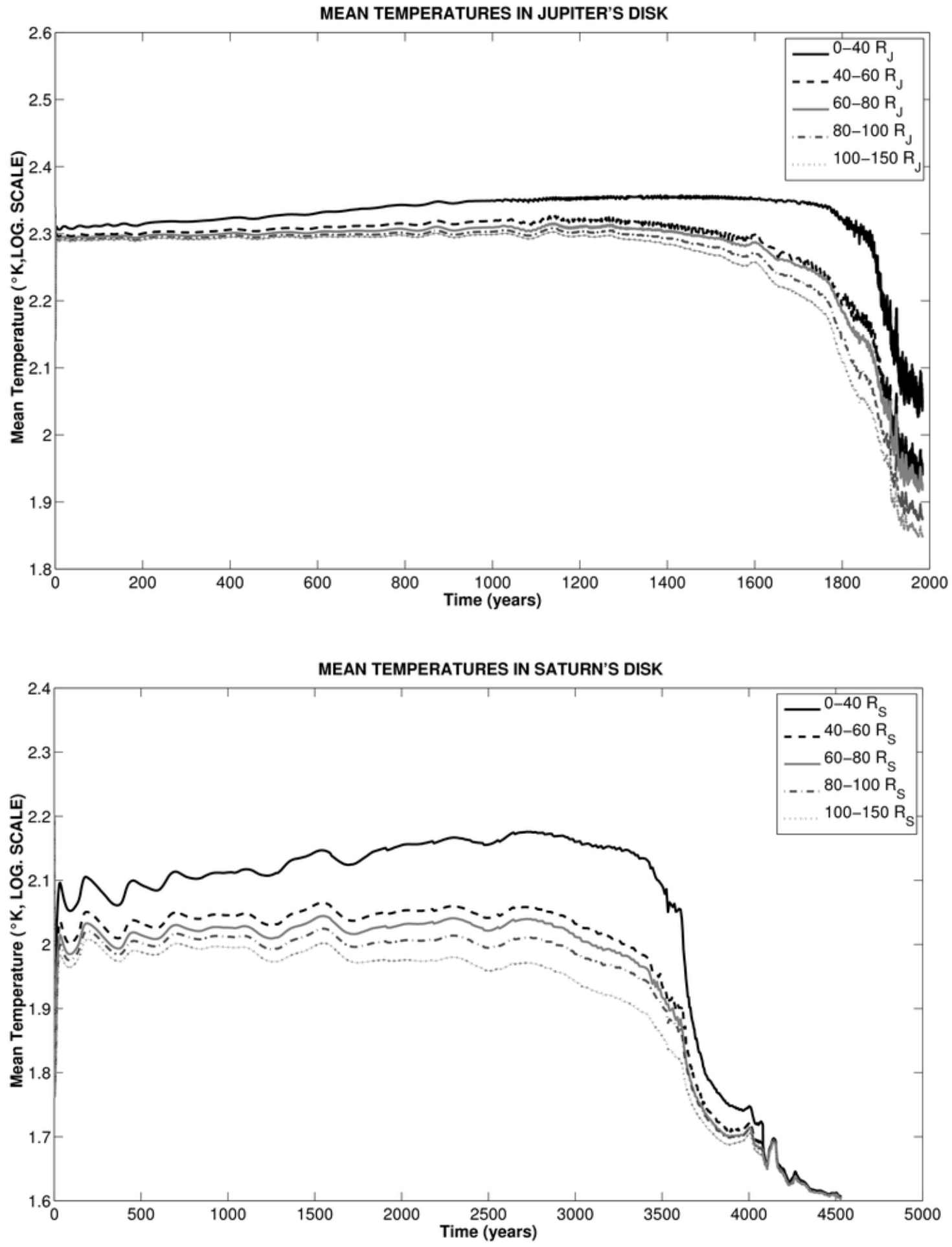

Figure 5: Mean temperatures in the Jovian and Saturnian disks during the formation of the respective planets. Each disk has been divided into five annular shells centered on the forming planets and the temperatures are averaged over each shell. Temperatures are expressed in °K and in logarithmic scale while the time is expressed in years.

*Figure 6*

**Jupiter**

**Panel a**

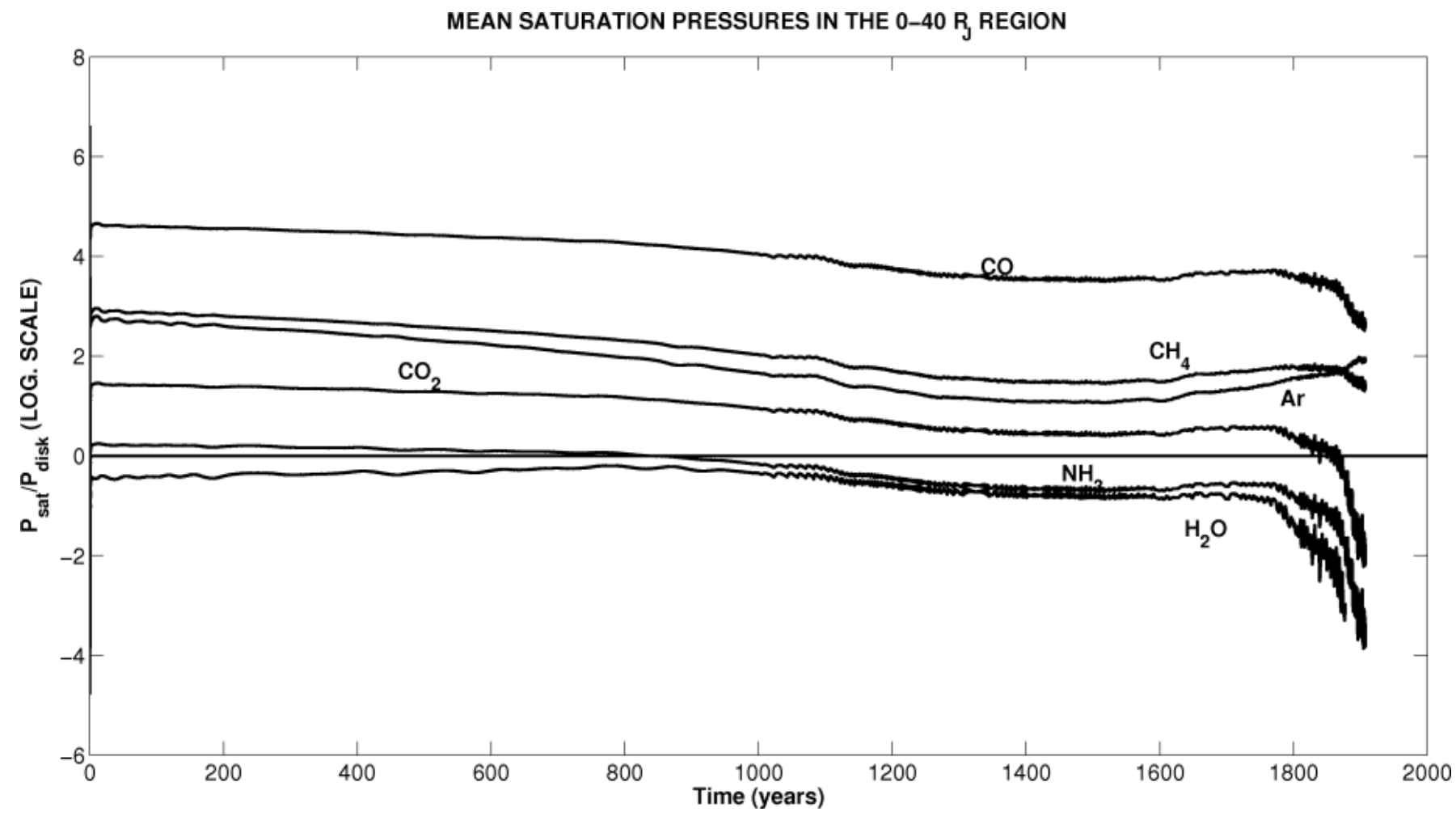

**Panel b**

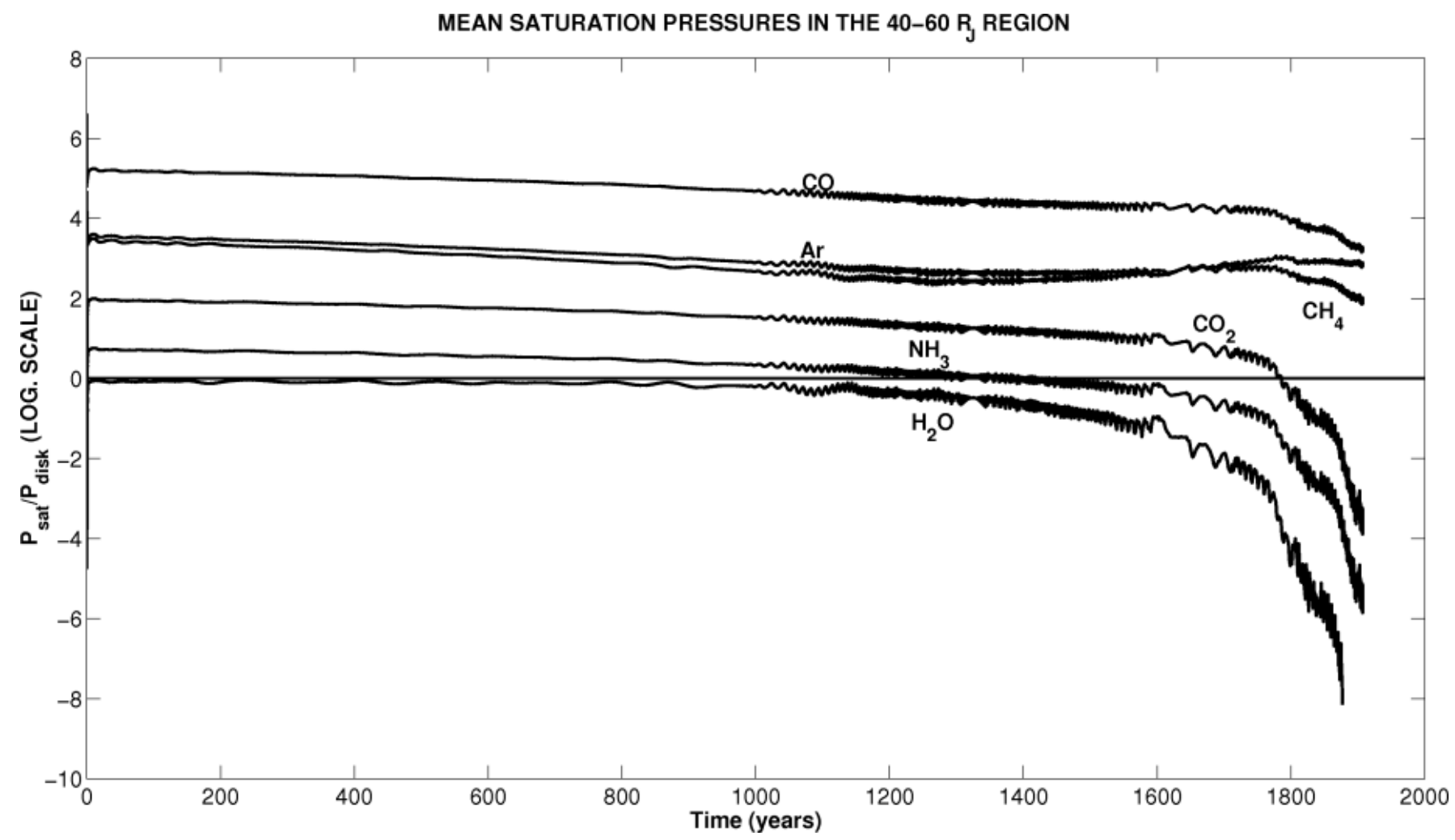

**Panel c**

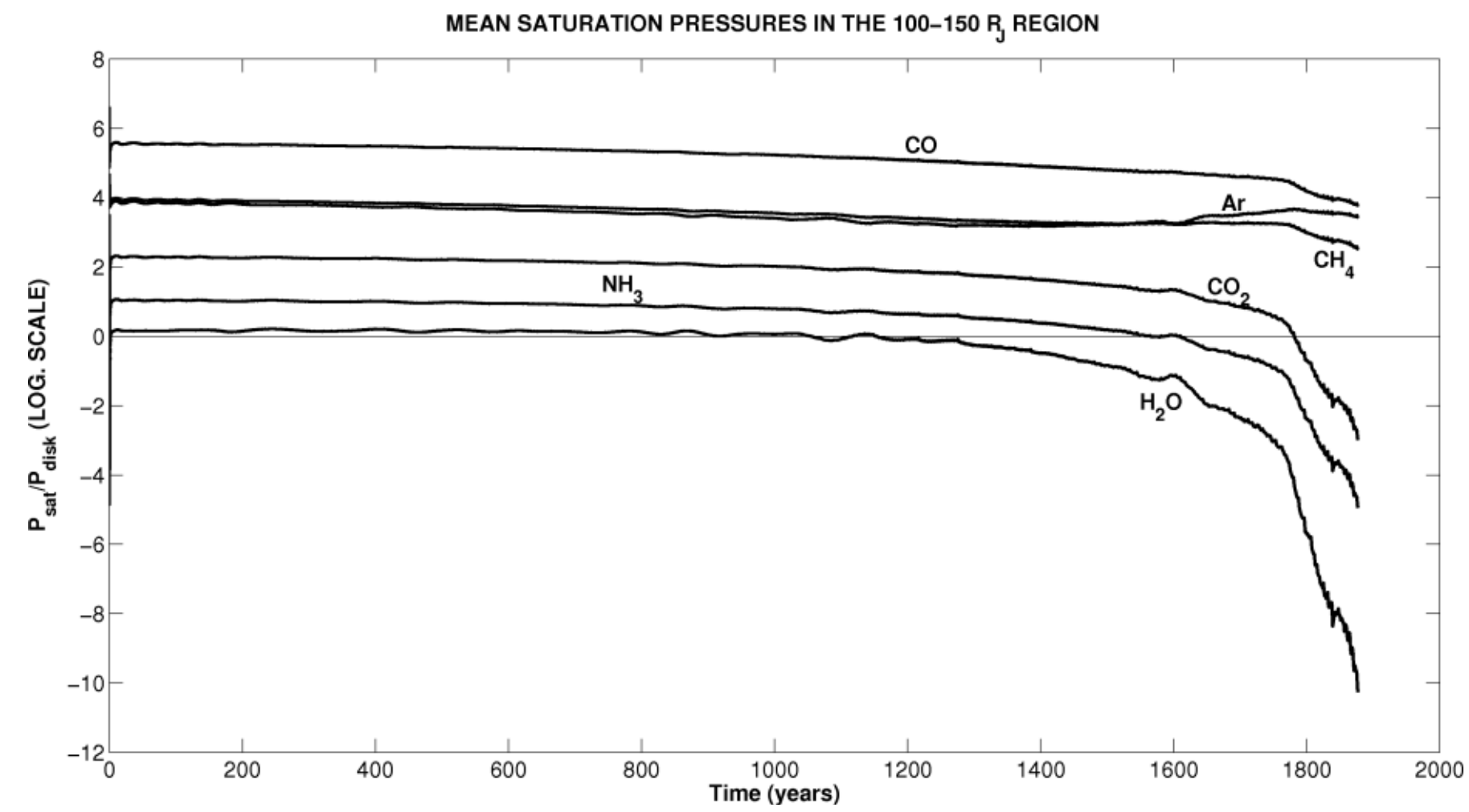

**Saturn**

**Panel d**

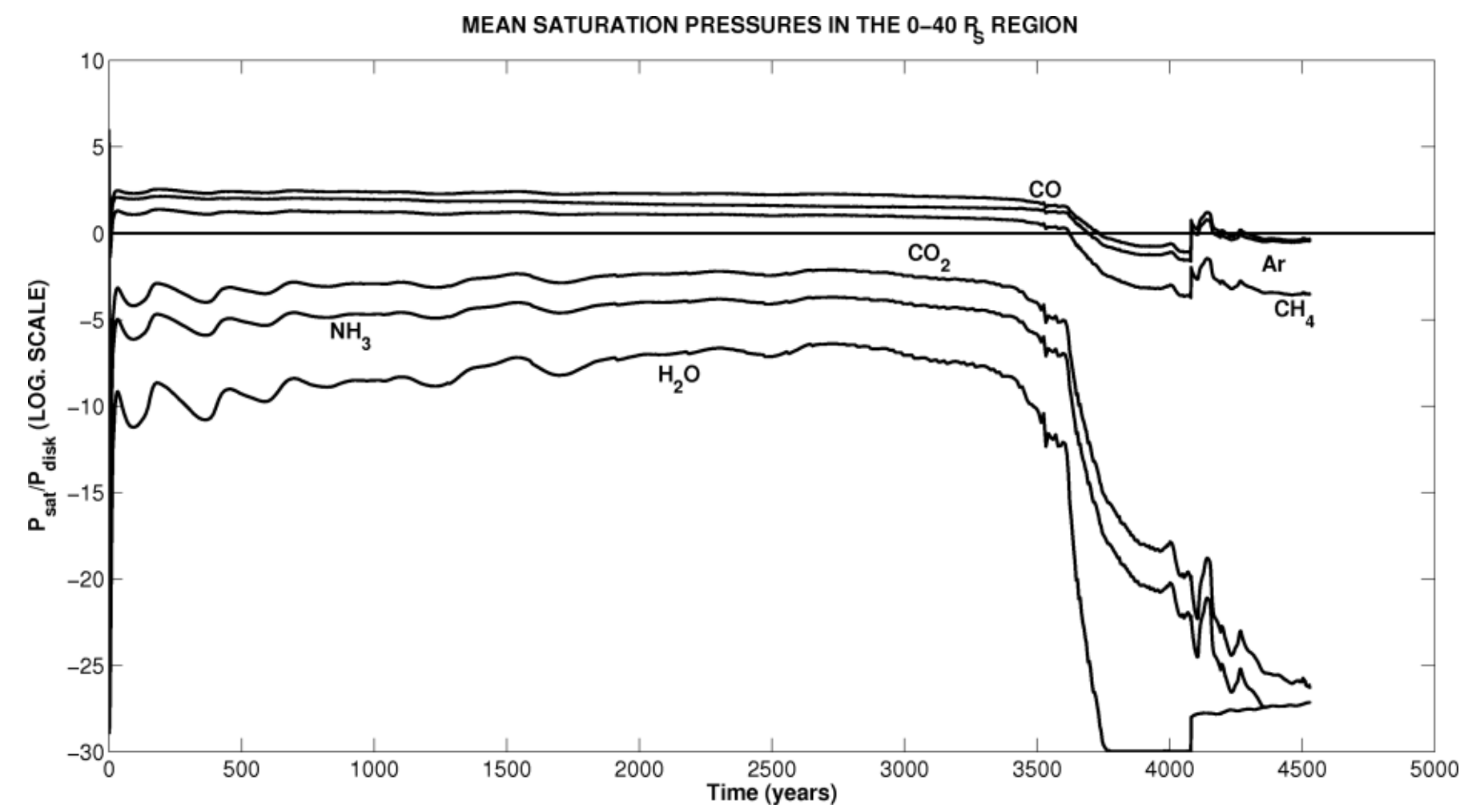

**Panel e**

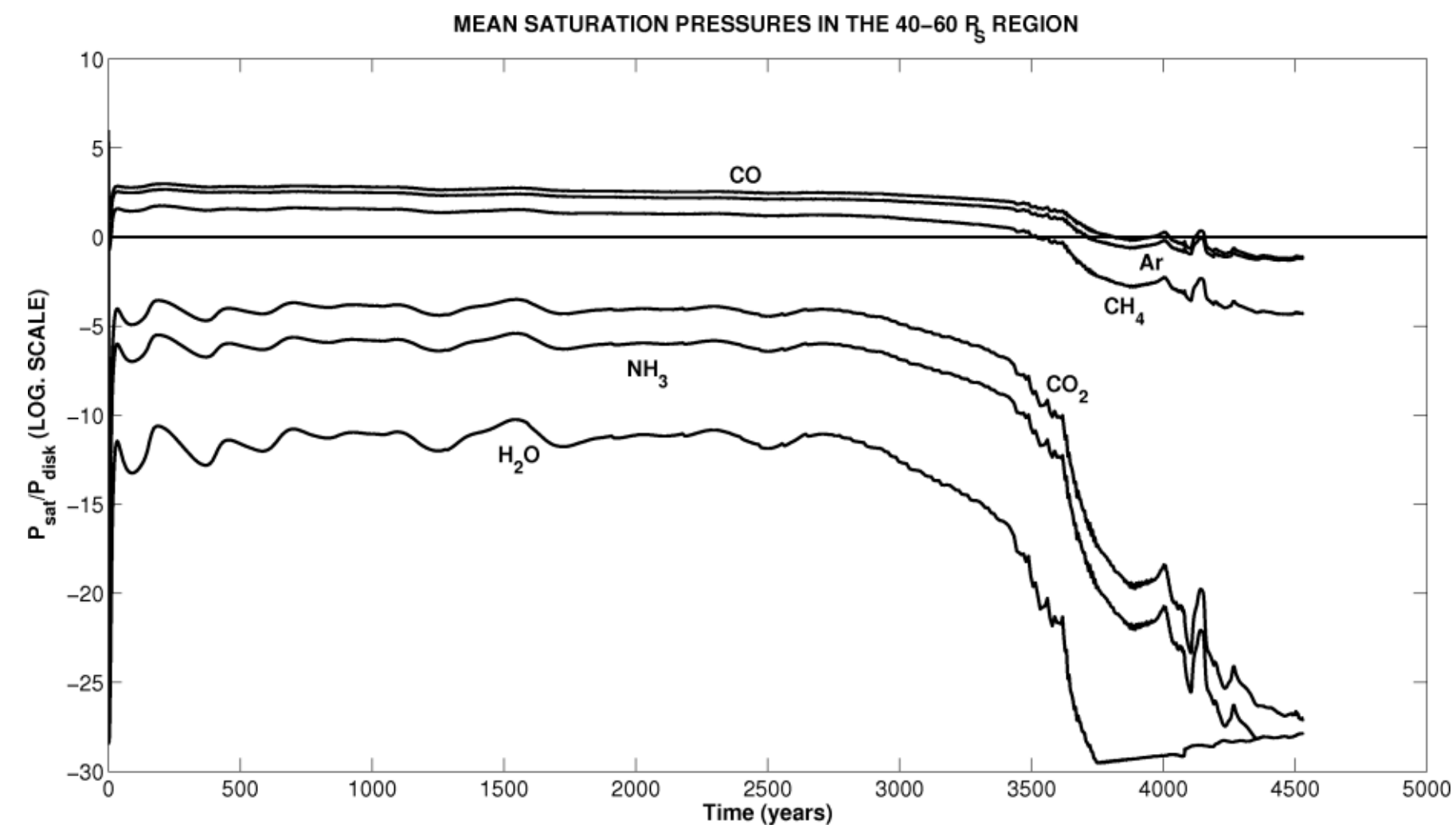

## Panel f

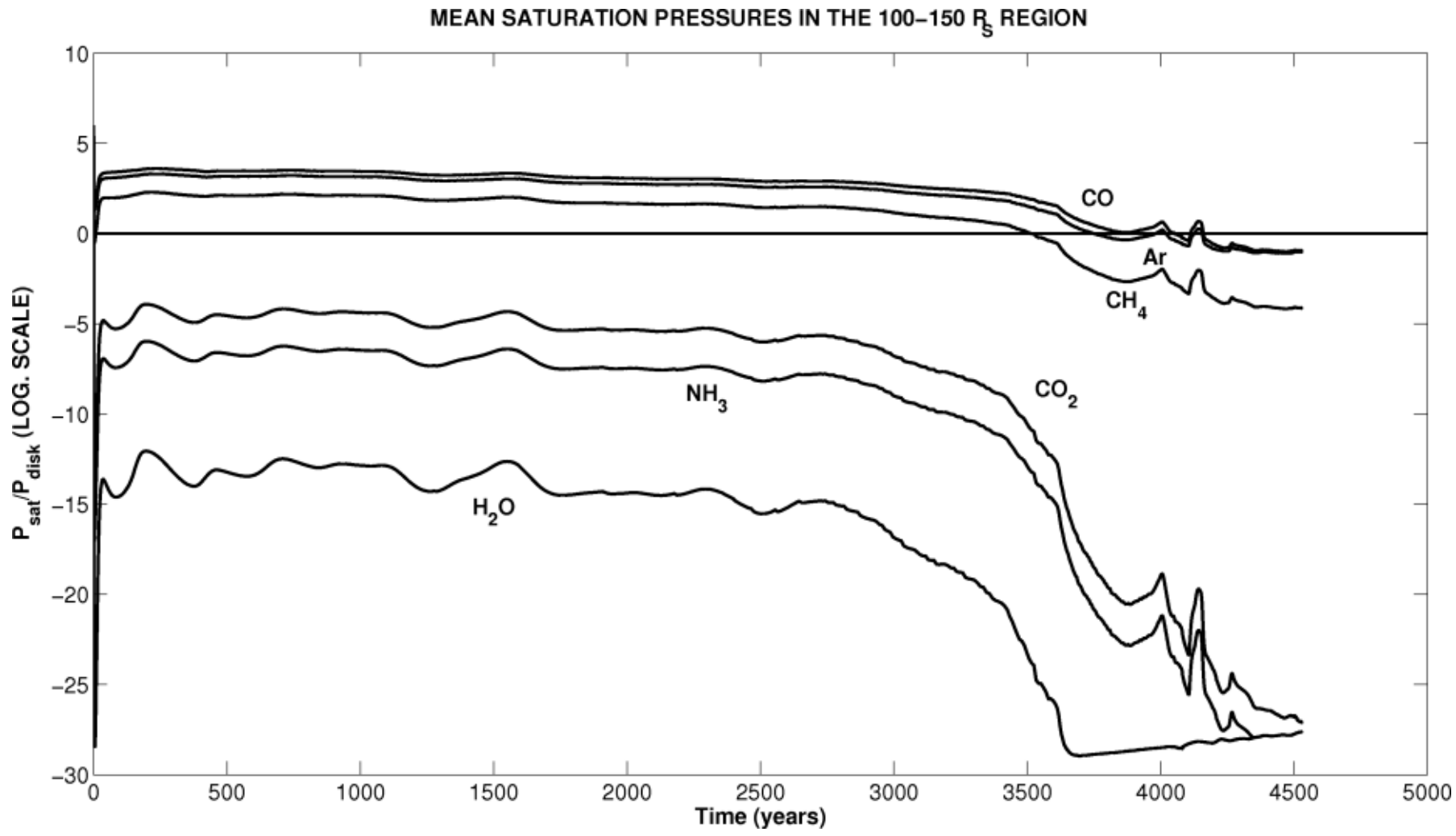

Figure 6: Mean vapor saturation pressures in the Jovian (panels a-c) and the Saturnian (panel d-f) disks as a function of time. Each panel shows the partial saturation pressures of different chemical species (CO, Ar, $CH_4$, $CO_2$, $NH_3$, $H_2O$) in selected annular shells as those described in Figure 5. The saturation pressures are expressed in units of the local pressure and in logarithmic scale while the time is expressed in years. As mentioned in the text, these figures do not represent an exact evaluation of the disk chemical evolution, since we do not compute which are the species that are present in the disk at each time step. Our calculations are indication whether provided certain species are present in the disk, the thermodynamic conditions could allow their condensation.